\documentclass[a4paper,11pt]{article}
\pdfoutput=1 

\usepackage{jcappub} 

\usepackage[T1]{fontenc} 
\usepackage{diagbox}
\usepackage{siunitx}
\usepackage{orcidlink}

\usepackage{lineno}

\title{\boldmath Directionality of low-energy electron recoils in the CYGNO experiment}

\author[a]{F.D. Amaro \orcidlink{0000-0001-7315-0550}}
\author[b,c]{E. Baracchini \orcidlink{0000-0003-4686-128X}}
\author[d]{L. Benussi \orcidlink{0000-0002-2363-8889}}
\author[d]{C. Capoccia \orcidlink{0009-0008-5919-3130}}
\author[d,e]{M. Caponero \orcidlink{0000-0002-5728-3123}}
\author[f]{L.G.M de Carvalho \orcidlink{0009-0003-5836-4771}}
\author[g,h]{G. Cavoto \orcidlink{0000-0003-2161-918X}}
\author[d]{I.A. Costa \orcidlink{0000-0002-3064-8305}}
\author[b,c]{M. D'Astolfo \orcidlink{0009-0000-9817-6693}}
\author[h]{G. D'Imperio \orcidlink{0000-0002-2945-0983}}
\author[d]{G. Dho \orcidlink{0000-0001-9454-9894}}
\author[h]{E. Di Marco \orcidlink{0000-0002-5920-2438}}
\author[a]{J.M.F. dos Santos \orcidlink{0000-0002-8841-6523}}
\author[b,c,1]{D. Fiorina \orcidlink{0000-0002-7104-257X}\note{Corresponding author}}
\author[h]{F. Iacoangeli \orcidlink{0000-0003-0808-5857}}
\author[i,b]{H. P. Lima Jr \orcidlink{0000-0001-7398-3237}}
\author[d]{G. Maccarrone \orcidlink{0000-0002-7234-9522}}
\author[j,k]{P. Meloni \orcidlink{0009-0001-7634-370X}}
\author[g,h]{A. Messina \orcidlink{0000-0003-1195-6780}}
\author[a]{C.M.B. Monteiro \orcidlink{0000-0002-1912-2804}}
\author[f]{R.A. Nobrega \orcidlink{0000-0001-5199-308X}}
\author[b,c]{G. M. Oppedisano \orcidlink{0009-0001-8801-2393}}
\author[f]{I.F. Pains \orcidlink{0009-0004-0851-6308}}
\author[d]{E. Paoletti}
\author[j,k]{F. Petrucci \orcidlink{0000-0002-5278-2206}}
\author[b,c]{S. Piacentini \orcidlink{0000-0002-1256-7149}}
\author[d]{D. Pierluigi}
\author[h]{D. Pinci \orcidlink{0000-0002-7224-9708}}
\author[d]{A. Russo}
\author[d,l]{G. Saviano}
\author[b,c]{F.F. Scamporlino \orcidlink{0009-0000-2937-993X}}
\author[a]{R.C. Sousa \orcidlink{0009-0000-4079-5213}}
\author[m]{N. J. C. Spooner}
\author[d]{R. Tesauro \orcidlink{0009-0006-0722-5896}}
\author[d]{S. Tomassini \orcidlink{0000-0001-7290-2028}}
\author[b,c,2]{S. Torelli \orcidlink{0000-0003-3622-3524}\note{Corresponding author, now at DIPC - Donostia International Physics Center}}

\affiliation[a]{LIBPhys; Department of Physics; University of Coimbra; 3004-516 Coimbra; Portugal}
\affiliation[b]{Gran Sasso Science Institute; 67100; L'Aquila; Italy}
\affiliation[c]{Istituto Nazionale di Fisica Nucleare; Laboratori Nazionali del Gran Sasso; 67100; Assergi; Italy}
\affiliation[d]{Istituto Nazionale di Fisica Nucleare; Laboratori Nazionali di Frascati; 00044; Frascati; Italy}
\affiliation[e]{ENEA Centro Ricerche Frascati; 00044; Frascati; Italy}
\affiliation[f]{Universidade Federal de Juiz de Fora; Faculdade de Engenharia; 36036-900; Juiz de Fora; MG; Brasil}
\affiliation[g]{Dipartimento di Fisica; Universit\`a di Roma Sapienza; 00185; Roma; Italy}
\affiliation[h]{Istituto Nazionale di Fisica Nucleare; Sezione di Roma; 00185; Roma; Italy}
\affiliation[i]{Centro Brasileiro de Pesquisas Físicas; Rio de Janeiro 22290-180; RJ; Brazil}
\affiliation[j]{Dipartimento di Matematica e Fisica; Universit\`a Roma TRE; 00146; Roma; Italy}
\affiliation[k]{Istituto Nazionale di Fisica Nucleare; Sezione di Roma Tre; 00146; Rome; Italy}
\affiliation[l]{Dipartimento di Ingegneria Chimica; Materiali e Ambiente; Sapienza Universit\`a di Roma; 00185; Roma; Italy}
\affiliation[m]{Department of Physics and Astronomy; University of Sheffield; Sheffield; S3 7RH; UK}

\emailAdd{davide.fiorina@gssi.it} 
\emailAdd{samuele.torelli@dipc.org}

\abstract{
Electron-recoil directionality can help distinguish solar-neutrino interactions from backgrounds and, together with recoil-energy measurements and three-dimensional tracking, enable neutrino-energy spectroscopy.
We study the reconstruction of the initial direction of low-energy electron recoils in the CYGNO optical time-projection chamber, focusing on the \SI{50}{\litre} LIME prototype and its high-granularity scientific CMOS (sCMOS) readout.
A dedicated algorithm identifies the interaction point from the asymmetry of the track light distribution and estimates the recoil direction from the early portion of the ionization trail, before multiple scattering erases most of the directional information.
The method is characterized with a detailed Monte Carlo simulation of electron recoils and detector response.
Samples spanning \SIrange{15}{70}{\kilo\electronvolt} are used to optimize the algorithm parameters by minimizing the width of the reconstructed-minus-true projected-angle distribution.
The projected angular response is modeled by two Gaussian components that account for its dependence on track inclination.
The narrower component, associated with favorable projections approximately parallel to the Gas Electron Multiplier (GEM) plane, has a width that improves from about \SI{20}{\degree} near \SI{20}{\kilo\electronvolt} to about \SI{12}{\degree} at \SI{60}{\kilo\electronvolt}.
We also quantify the fraction of events in the correctly reconstructed angular core and discuss the dominant limitations of a camera-only two-dimensional measurement.
}

\begin{document}
\maketitle
\flushbottom

\section{Introduction}
\label{sec:intro}

Reconstructing the initial direction of a recoil offers several advantages in physics scenarios where signal events are expected to have a preferred spatial direction. Directional information can serve as an additional tool for background rejection by considering events originating from interactions with a direction compatible with the expected signal.

An important application of electron-recoil directionality is solar-neutrino detection through elastic neutrino--electron scattering. The correlation between the recoil direction and the known solar direction provides a distinctive signature, allowing solar-neutrino interactions to be statistically separated from approximately isotropic backgrounds. Moreover, measuring the electron-recoil energy and its three-dimensional scattering angle allows the incident neutrino energy to be inferred from elastic-scattering kinematics, opening the possibility of solar-neutrino spectroscopy \cite{Lisotti2024}. This capability complements the directional dark-matter programme of gaseous TPCs, which relies primarily on nuclear recoils.
Electron directionality is also central to photoelectric
X-ray polarimetry, where the azimuthal distribution of
photoelectron emission encodes the polarization of the
incident radiation. The extension of this technique to
hard X-rays using optically read out triple-GEM TPCs has
recently been investigated \cite{Fiorina2025Polarimetry}.

However, measuring the initial direction of low-energy electron recoils in a time-projection chamber (TPC) presents a significant challenge due to the combined effects of multiple scattering of the recoil electron in the gas and diffusion of the ionization electrons produced along its track. Multiple scattering progressively erases information on the initial recoil direction, while diffusion reduces the accuracy with which the interaction point can be determined.

The CYGNO project aims to develop a gaseous high-precision TPC with an optical readout for directional Dark Matter searches and solar neutrino spectroscopy \cite{Lisotti2024}. CYGNO incorporates innovative features, such as the utilization of a He-CF$_4$ scintillating gas mixture at atmospheric pressure and room temperature, together with a triple gas electron multiplier (GEM)
amplification system \cite{SAULI1997531},
optically read out by photomultiplier tubes (PMTs)
and scientific CMOS (sCMOS) cameras. By combining the information of the high-granularity camera with the fast sampling of the PMT, it is possible to perform 3D tracking with head-tail capability and particle identification down to O(keV) energy \cite{Amaro:2025BayesianNetworkCygno}. The latest detector developed by the collaboration is the LIME \cite{CYGNO:2023ucc} prototype with a 33x33 cm$^2$ readout area and a drift length of 50 cm, for a total of 50 L active volume, and it is read out by 1 sCMOS camera and 4 PMTs.

To assess the performance of the initial direction measurement of electron recoils in the CYGNO gas TPC, a dedicated algorithm has been developed and characterized, and the angular resolution achievable with the detector has been studied. Due to the challenges of finding a source of low-energy electron recoils with a known direction, the directionality studies have been carried out on LIME simulated tracks generated using the simulation outlined in \cite{Amaro:2026trz}. This simulation demonstrates a strong agreement with experimental data, particularly in terms of light response and energy resolution, but also in terms of the reproduction of the track shape variables.

In this paper, after an introduction on the detector and the event reconstruction (Sec. \ref{sec:Det}), we present a directionality reconstruction algorithm used to determine the initial direction of recoil tracks in a LIME-like optical TPC (Sec. \ref{sec:dirAlgo}). The procedure begins by identifying the start of the track, i.e., the interaction point, through the determination of the track extremities and the exploitation of the characteristic asymmetry in the energy–loss profile. Once the interaction point is established, the recoil direction is inferred by following the energy-deposit distribution from this point while restricting the analysis to the early portion of the track, where directional information remains preserved before being degraded by multiple scattering.

The directionality algorithm has been adapted, optimized, and validated using a dedicated sample of Monte Carlo–simulated electron recoils produced in a LIME-like detector configuration (Sec. \ref{sec:sim}). For each simulated event, the full pixel-level light information—namely the measured intensity associated with each pixel belonging to the reconstructed track together with its two-dimensional position in the image—is used as input to the reconstruction. This dataset has been used specifically for the optimization of the algorithm parameters (Sec. \ref{sec:paroptim}) and provides a direct comparison with Monte Carlo truth, allowing for the evaluation of the reconstruction performance (Sec. \ref{sec:perform}). In the last section, the efficiency of the directionality algorithm is assessed and discussed (Sec. \ref{sec:AlgoEff}).
\section{LIME detector and event reconstruction}
\label{sec:Det}
\begin{figure}
    \centering
    \includegraphics[width=0.9\linewidth]{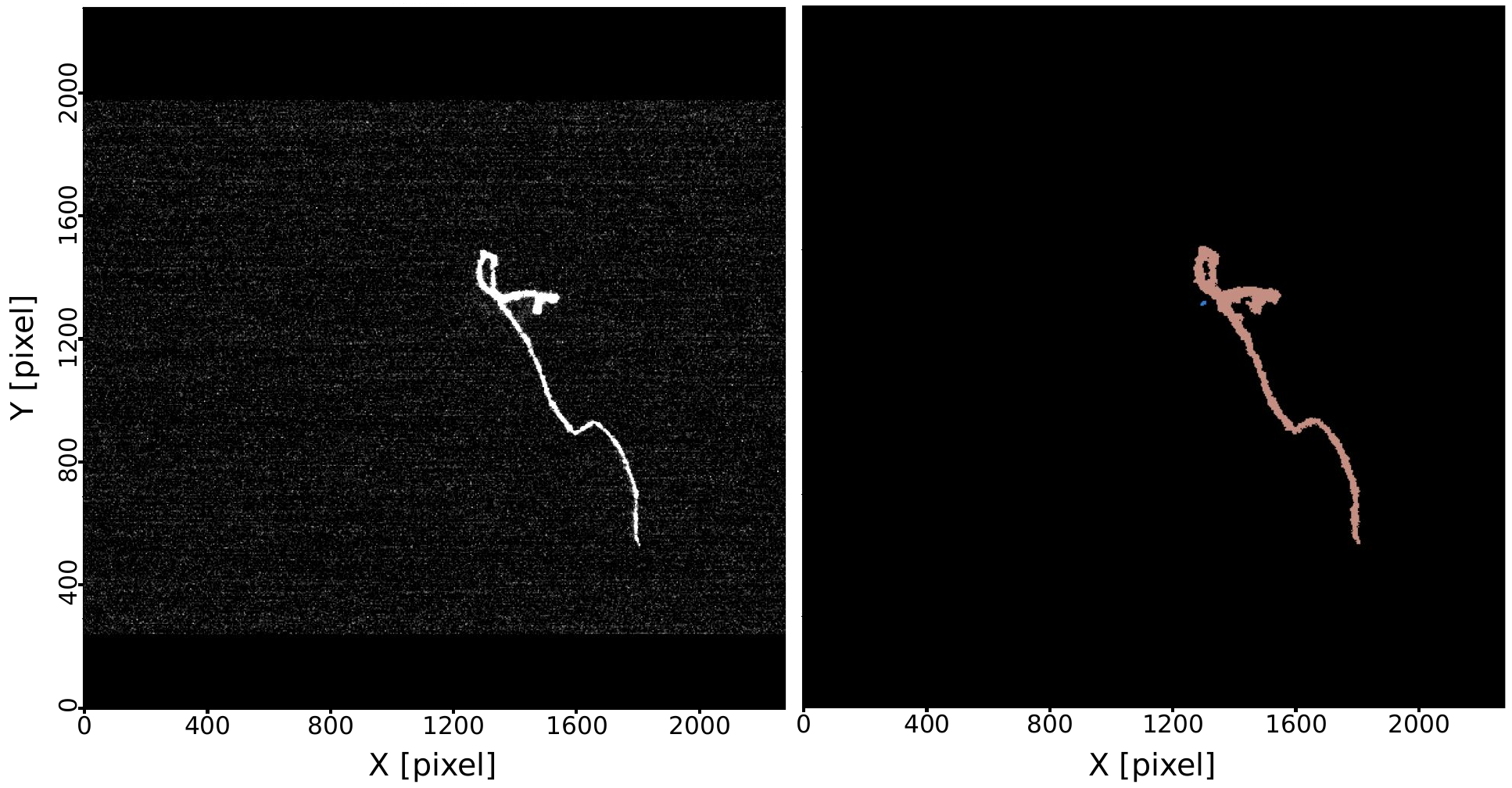}
    \caption{Left: Original sCMOS image of an electron recoil acquired with the LIME detector underground. Right: reconstructed tracks; the coloured clusters represent the electron tracks.}
    \label{fig:traccine}
\end{figure}

LIME~\cite{CYGNO:2023ucc} is a gaseous Time Projection Chamber prototype 
developed within the CYGNO program to study low-energy nuclear recoils using 
optical readout. The detector operates with a He:CF$_4$ (60:40) gas mixture at 
atmospheric pressure and features a drift region coupled to a triple Gas Electron 
Multiplier (GEM) amplification stage. 
The low gas density produces extended recoil tracks,
while the high granularity of the optical readout
allows detailed sampling of their topology.
Ionization electrons produced by charged 
particles drift toward the GEMs, where charge amplification occurs and 
scintillation light is emitted in the gas. 
The scintillation light generated during the avalanche
process is collected by a Hamamatsu ORCA-Fusion
scientific CMOS (sCMOS) camera \cite{CYGNO:2023ucc}.
Each acquired frame records the two-dimensional
distribution of the scintillation light emitted in
the GEM amplification stage on a $2304\times2304$
pixel sensor. The physical pixel pitch is
$6.5\,\mu\mathrm{m}$; with the optical magnification
used, each pixel corresponds to an area of
approximately $152\times152\,\mu\mathrm{m}^2$
on the GEM plane.
Low-energy 
electron recoils appear as faint and short tracks, composed of groups of pixels 
with varying intensity, reflecting the local ionization density combined with the 
effects of diffusion during drift.
Data reconstruction \cite{Baracchini_2020} begins with raw sCMOS images. After pedestal subtraction and noise suppression to retain only significant pixels, the images are corrected for optical non-uniformities. The signal pixels selected are grouped into clusters using a variant of the Density-Based Spatial Clustering of Applications with Noise (DBSCAN) algorithm described here \cite{Baracchini_2020}, where each resulting cluster corresponds to a single recoil track. The final reconstructed track is therefore described as a set of pixels, each defined by its spatial coordinates (x,y) on the sensor and by an associated intensity proportional to the detected light yield. From this information, the total light, geometrical properties, and intensity-weighted moments of the pixel distribution are computed, forming the basis for energy, topology, and directional analyses. An example of sCMOS image with associated reconstruction of the pixels associated to tracks is shown in Fig. \ref{fig:traccine}

\section{Low energy electron recoils track direction reconstruction algorithm}
\label{sec:dirAlgo}
\begin{figure}
    \centering
    \includegraphics[scale=.45]{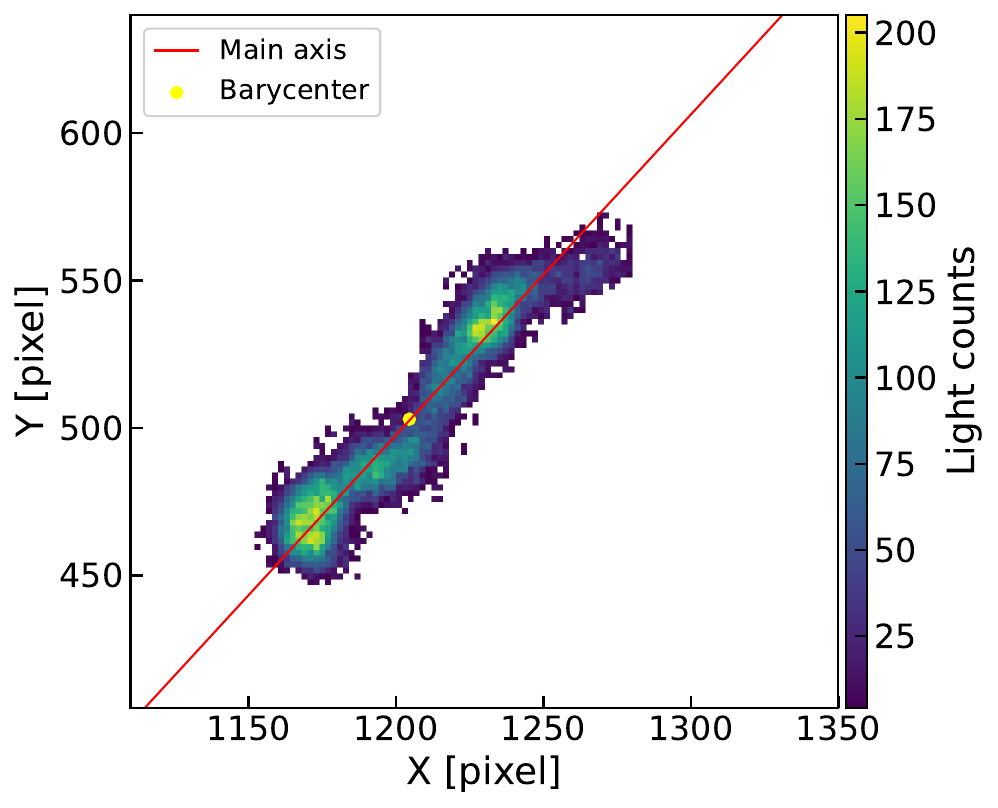}
    \includegraphics[scale=.45]{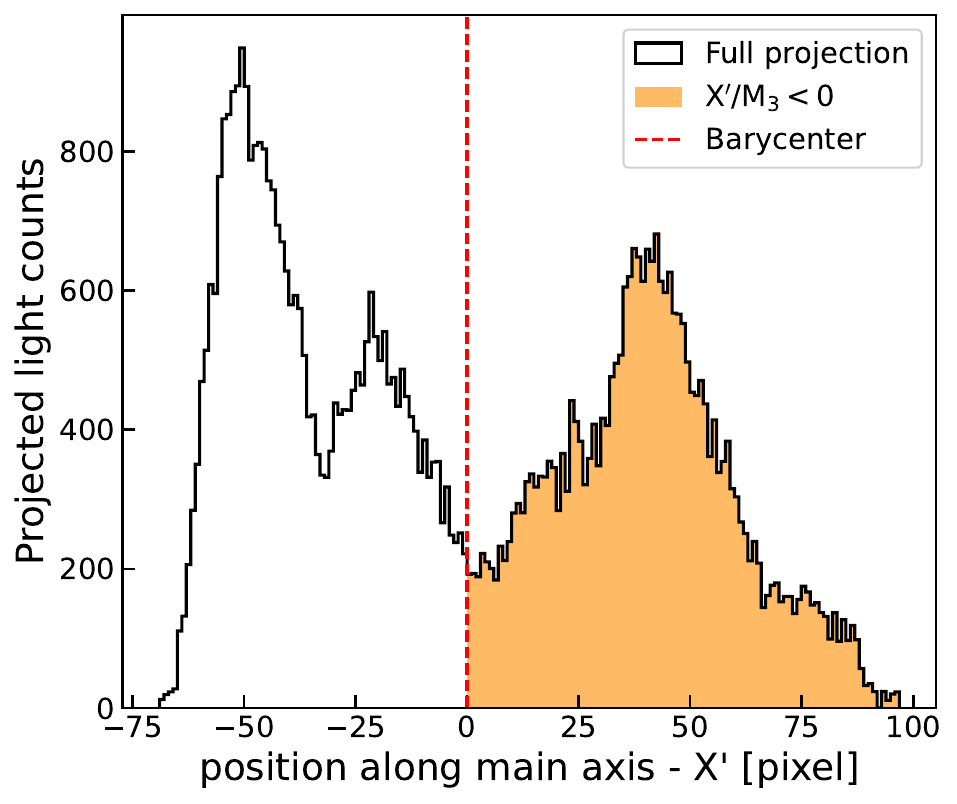}
    \caption{Left: simulated 40\,keV electron recoil with its barycenter and main 
    axis calculated through the procedure described in this section. Right: Charge distribution projected onto the main axis relative to the barycenter (origin). The orange shaded area represents the region opposite the Bragg peak, identified by the selection criteria specified in the legend.}
    \label{fig:trackBarMa}
\end{figure}

The algorithm developed for this work has been adapted and optimized starting 
from a strategy originally developed for X-ray polarimetry~\cite{AstroXRayPol}, 
previously implemented for track reconstruction in teh Gas Pixel Detector (GPD)~\cite{BALDINI2021102628}. The Gas Pixel Detectors developed for X-ray polarimetry
are optimized for the 2--8 keV energy band
\cite{BALDINI2021102628}.
In contrast, the larger TPCs envisioned for CYGNO
are intended to investigate electron recoils over
a broader energy range, resulting in longer and more
varied track topologies. Therefore, the algorithm
has been adapted to handle this wider range of tracks.\\
The process starts with the determination of the track charge barycenter and 
the calculation of the track main axis passing through the barycenter, as shown in Fig.~\ref{fig:trackBarMa} (left). The main axis is determined by maximizing the standard deviation of the charge distribution, which is obtained by projecting the track charge onto a generic axis passing through the barycenter. The axis that yields this maximum is defined as the main axis.
In the energy range considered in this work (15--70\,keV), electron recoils 
exhibit a characteristic asymmetry in their energy-loss profile: as the 
particle slows down, the energy loss per unit length increases, leading to a 
significant concentration of energy deposition toward the end of the trajectory. 
As a result, electron recoil tracks present a tail with low ionization density 
and a head with higher ionization density, associated with the Bragg peak. 
This asymmetry, clearly visible in the track images at all energies considered 
and increasingly pronounced at lower energies where the Bragg peak is more 
localized, is the key feature exploited by the algorithm to determine the 
initial direction of the recoil. 
This asymmetry in energy release can be quantified by analyzing the third moment of the distribution along the track's main axis relative to the barycenter, commonly referred to as skewness (M$_3$).
The skewness enables the identification of points located on the opposite side of the Bragg peak relatively to the axis passing through the barycenter and perpendicular to the previously defined track's major axis.
Thus the algorithm follows with the determination of the interaction point (IP)
from this set as the point located at the extremity of this region with respect 
to the barycenter. To do so, a circle with radius $r$ centered at the barycenter 
is considered, and the points that satisfy the following two conditions are 
selected:
\begin{itemize}
    \item The point lies on the opposite side of the Bragg peak, respecting the condition X'/M$_3$<0, where X' is the coordinate of the point along the main axis;
    \item The point must be at a distance $d_\mathrm{cm} > r$ from the barycenter.
\end{itemize}
The procedure is repeated starting with a small value of $r$, incrementing it 
at each step, until a predetermined number of pixels, $\mathrm{N}_{pt}$, remains 
in the selection. The quantity $\mathrm{N}_{pt}$ is a model parameter, which is 
not known a priori, and its optimization will be discussed subsequently. The 
barycenter of the region consisting of the pixels that pass the selection is 
defined as the track interaction point.

\begin{figure}
    \centering
    \includegraphics[scale=.6]{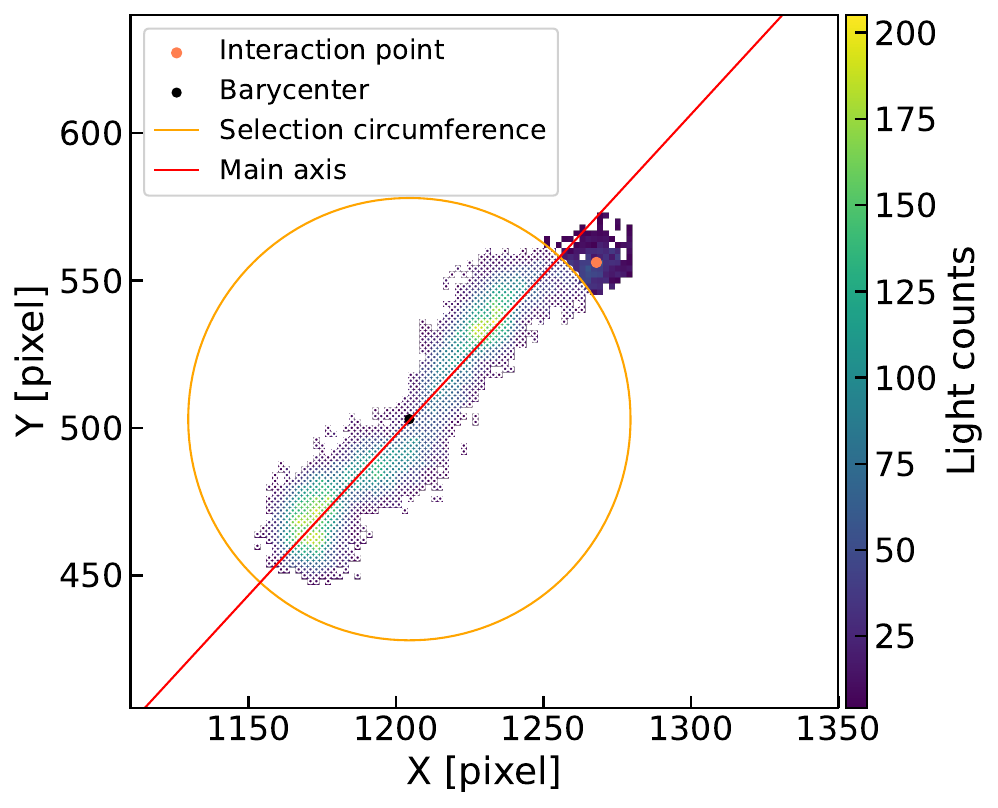}
    \caption{40 keV electron with main axis, barycenter, and selection circumference shown. The region outside the circumference in clear is the one selected for the interaction point calculation. In that region the interaction point is also shown.}
    \label{fig:IPCal}
\end{figure}

It should be noted that this procedure is best suited for tracks with a 
relatively simple topology. For highly curved or complex tracks — such as those 
produced at lower energies where multiple scattering is more severe — the 
algorithm may fail to correctly identify the interaction point. In such cases, 
the expanding-circle approach can inadvertently exclude pixels near the true 
interaction point if these happen to lie close to the barycenter, or 
erroneously select pixels from an unrelated portion of the track. These failure 
modes contribute to the flat component observed in the $\theta_\mathrm{meas} - 
\theta_\mathrm{true}$ distribution and are 
discussed in detail later in the text.
This selection procedure is represented in Fig. \ref{fig:IPCal}.
The interaction point is calculated using a weighted average, rather than relying on the track's extremity, due to the transverse diffusion of electrons originating from the actual interaction point. As ionization electrons diffuse during drift, they spread out in a cloud around it. Consequently, determining the interaction point as the weighted average of the initial portion of the track has been shown to be an effective approach for its identification \cite{AstroXRayPol}.
Once the interaction point of the track is identified, the track direction is determined by analyzing the light intensity along the track profile. However, the directional information is primarily contained at the start of the track and diminishes progressively along its length. Therefore, the intensity of each pixel along the track is adjusted by the following factor:

\begin{equation}
W(d_{ip}) = e^{-d_{ip}/\mathrm{w}}
\end{equation}

where $d_{ip}$ is the distance between the pixel and the interaction point, and $w$ is the characteristic distance scale, expressed in pixels, controlling the exponential weighting, another free parameter of the algorithm, whose optimization is discussed later. This approach leads to an exponential decay in the weighting of the pixels as the distance from the interaction point increases, thereby giving greater importance to pixels near the IP. The track direction $\theta$ is then calculated as the main axis of the track passing through the interaction point. The procedure described so far enables the determination of the track's direction, but not its orientation. The orientation is then defined by following the light path along the track close to the IP. To assign an orientation to the track, the barycenter of the IP region is calculated using a smaller radius $r$ and the position of this barycenter is used to give an orientation to the track direction. By selecting a larger number of pixels, the barycenter of the selected points follows the light pattern along the track. Therefore, this information can be used to establish the track's orientation.
\begin{figure}
    \centering
    \includegraphics[scale=.6]{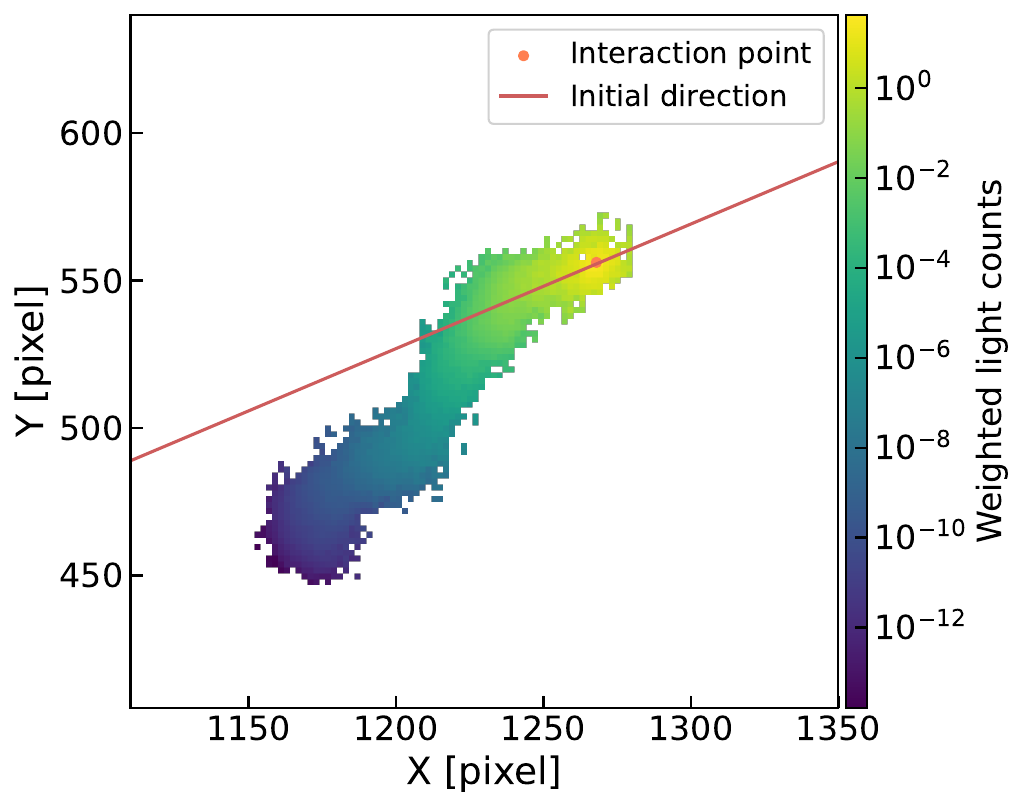}
    \caption{Plot of the rescaled track, the interaction point, and the line of the direction of the track is shown.}
    \label{fig:RescaledOri}
\end{figure}
The orientation is then given following the light path  from the interaction point.
\section{Electron recoil directionality performance}

\subsection{Simulation of sCMOS images of low-energy electron recoil}
\label{sec:sim}
The simulation of recoil tracks begins with the generation of the physical ionization trail using GEANT4 \cite{GEANT4:2002zbu}, which provides the three-dimensional pattern of energy deposits along the particle trajectory. From this input, the detector response is modeled through a digitization algorithm that reproduces, in detail, the sequence of processes occurring inside the gaseous material and in the sCMOS sensor \cite{Amaro:2026trz}.
At each interaction point the simulation samples the number of primary ionization electrons according to Poissonian statistics, accounts for the probability of electron absorption in the gas during drift, and propagates the surviving charge toward the amplification region while accounting for longitudinal and transverse diffusion calibrated from experimental measurements. When the electrons enter the GEM stack, the simulation applies a multi-stage amplification model in which the first GEM includes stochastic gain fluctuations drawn from an exponential distribution, while subsequent stages use mean gains and include extraction inefficiencies, additional diffusion within the GEM holes, and the onset of local saturation in regions of high charge density \cite{Amaro2026LightResponse}. The avalanche electrons produce scintillation light, whose yield fluctuates according to the stochastic nature of photon production. This light distribution is then projected through the optical system, taking into account optical effects such as lens distortions, vignetting, and solid angle, before being digitized by a detailed model of the sCMOS camera that includes sensor noise and pixel response non-uniformity. Through extensive parameter tuning of diffusion, gain, saturation, and optical parameters, the algorithm produces simulated images that resemble LIME data with a good agreement in light response, energy resolution, and track topology as shown in \cite{Amaro:2026trz}.

\subsection{Parameters optimization}
\label{sec:paroptim}
The algorithm relies on two free parameters, $\mathrm{N}_{pt}$ and $\mathrm{w}$. Since the model used is empiric, the values of these parameters are not known a priori, and they are tuned with the target of the best angular resolution for tracks of our interest in a LIME-like detector with the algorithm described in \cite{Amaro:2023ppa}. Electron-recoil samples were generated at energies between 15 and 70 keV, with a spacing of 2 keV and 10,000 events per energy point. The algorithm parameters were optimized at the representative energies listed in Tab.~\ref{tab:parametersvsE}. The simulated recoil electrons have isotropic initial directions and uniformly distributed starting positions in the transverse plane, with drift distances uniformly distributed between 5 and 45 cm. The tracks were reconstructed using the code outlined in \cite{CYGNO:2023ucc}, with the same parameters applied as in the data. 
The directionality algorithm uses only the two-dimensional track image and is therefore sensitive to the initial track inclination relative to the GEM plane. Tracks approximately parallel to the plane retain most of their directional information in the projection. As the inclination increases towards the perpendicular direction, the projected initial segment becomes shorter and its direction is more difficult to reconstruct \cite{KITAGUCHI2018188}.
Since the simulated electron directions are isotropic, the inclusive angular response combines contributions from a continuous range of inclinations. For events retaining directional information, this response can be described as a superposition of approximately gaussian distributions with inclination-dependent widths. For simplicity, we approximate this continuous mixture with two gaussian components. The narrower component characterizes the more favourable projections, associated mainly with tracks approximately parallel to the GEM plane, while the broader component accounts for projections more strongly degraded by inclination. These components provide an effective description of the continuous response rather than identifying two distinct classes of track inclination.

\begin{figure}
    \centering
    \includegraphics[width=0.7\linewidth]{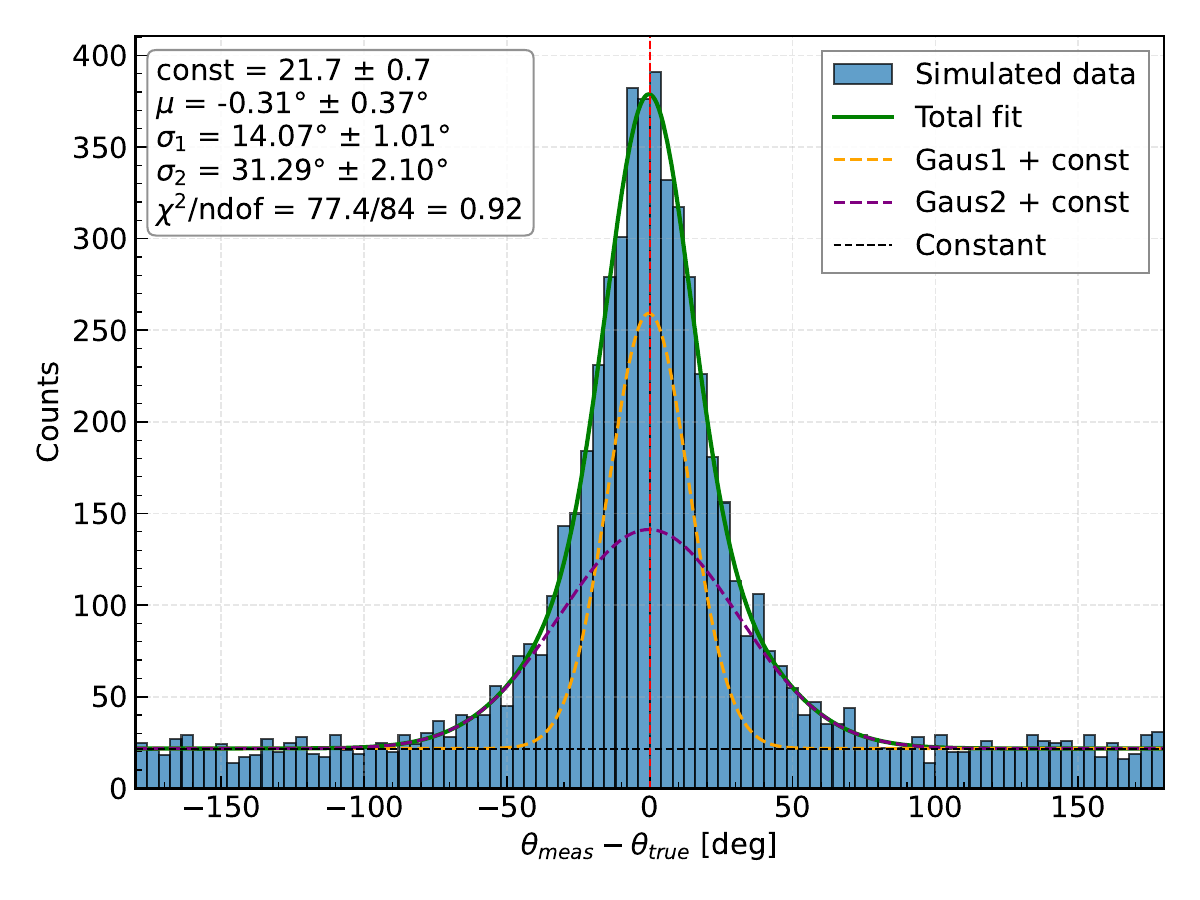}
    \caption{Distribution of $\mathrm{\theta}_{\mathrm{meas}}-\mathrm{\theta}_{\mathrm{true}}$ for 40 keV ER with fit consisting of two gaussian plus a constant term superimposed. The parameters in the legend are, respectively, the constant value, the mean of the two gaussians, the sigma of the two gaussians in degrees and the reduced $\chi^2$.}
    \label{fig:gaussiandistrib}
\end{figure}

Additionally, because of the methodology, the algorithm may struggle with complex track topology or those tracks that have an overlap in correspondence of the interaction point, leading to the misidentification of the interaction point. In these cases, the track direction reconstruction fails, resulting in a random direction as output. This contribution is modeled by a constant term, which is excluded from the quoted gaussian widths and accounted for separately through the directional reconstruction efficiency. The full fit function is therefore composed of two gaussians, with common mean, plus a constant term. The distribution of $\mathrm{\theta}_{\mathrm{meas}}-\mathrm{\theta}_{\mathrm{true}}$ for 40 keV electron recoil, with the fit superimposed, is shown in Fig. \ref{fig:gaussiandistrib}.

To optimize the algorithm free parameters, a systematic scan of the RMS of the distribution after subtracting the constant term and flat term amplitude was performed by varying $\mathrm{N}_{pt}$ and $\mathrm{w}$. The parameters tested include all combinations of $\mathrm{N}_{pt}$ from 60 to 200 in steps of 10 and $\mathrm{w}$ from 1 to 10 in steps of 0.5. These bounds define the empirical range explored for the two reconstruction parameters and do not represent physical thresholds. For the optimization, the RMS of the whole distribution after the subtraction of the constant term, has been considered. The optimization targets the angular RMS of the directional
component. The fitted constant contribution was also monitored
and found to vary only weakly near the minimum. The directional
reconstruction efficiency and the main failure mechanisms
are discussed separately in Sec.~\ref{sec:AlgoEff}.

\begin{figure}
    \centering
    \includegraphics[width=.8\linewidth]{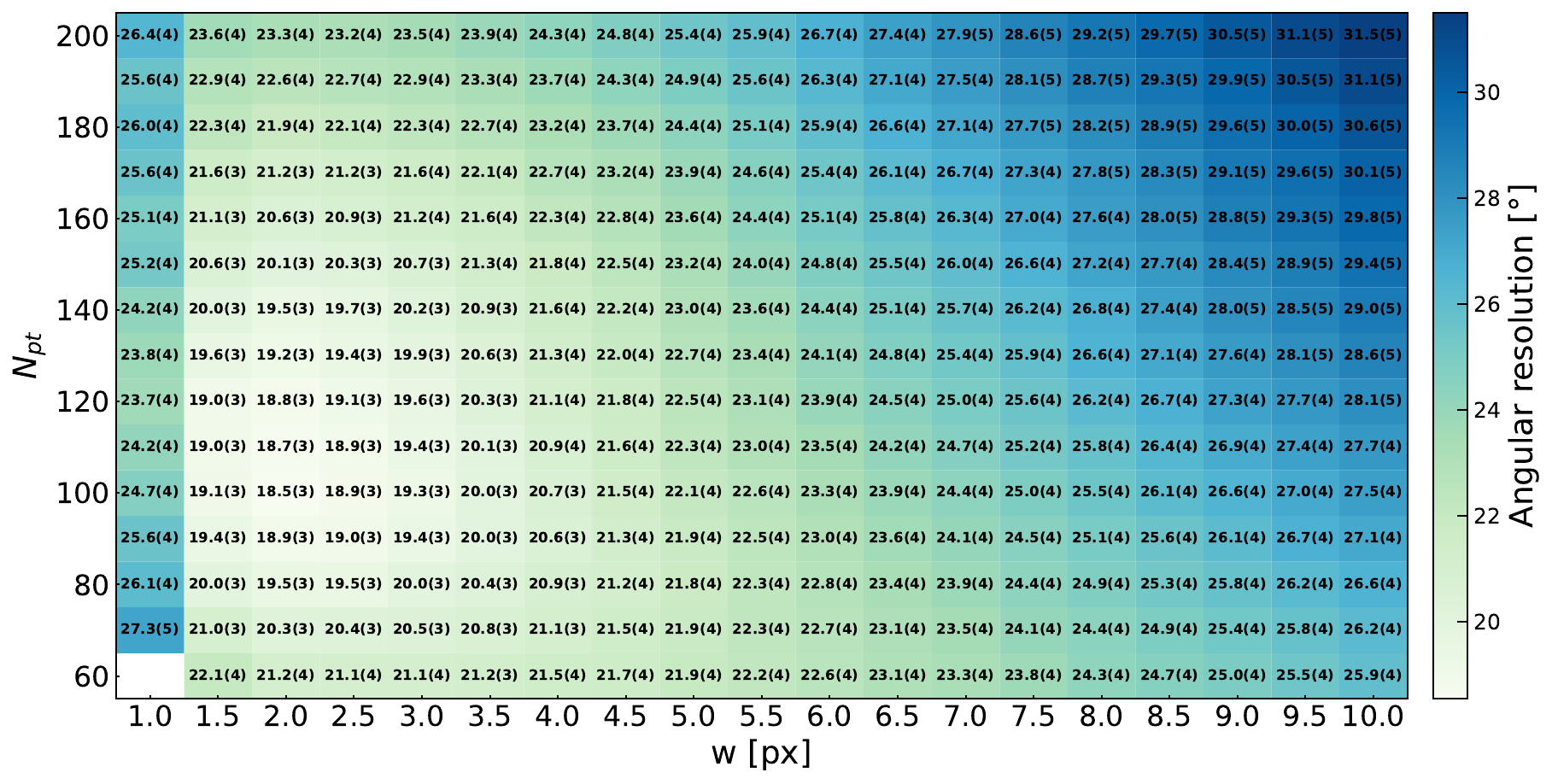}
    \caption{Angular resolution obtained for the 40 keV dataset as a function of the two parameters of the algorithm $\mathrm{N}_{pt}$ and $\mathrm{w}$. In each bin, the numerical value of angular resolution is reported as the $RMS_{\theta}$ of the $\theta-\mathrm{\theta}_{\mathrm{true}}$ distribution after subtraction of the constant term, together with the error between parenthesis referring to the statistical error on the RMS. The color scale represents the angular RMS,
with smaller values corresponding to better angular resolution. }
    \label{fig:AngResVsPar}
\end{figure}
For each value of $\mathrm{N}_{pt}$, $RMS_{\theta}$ improves rapidly as $\mathrm{w}$ increases, reaches a minimum, and then begins to rise again. This behavior reflects the role of the $\mathrm{w}$ parameter. When $\mathrm{w}$ is too small, a portion of the track that is overly short and dominated by diffusion is used to evaluate the direction, leading to no directional information. As $\mathrm{w}$ increases, the length of the track considered for direction evaluation expands, enhancing the sensitivity to directional information until an optimal value is reached. However, further increasing $\mathrm{w}$ brings in segments of the track affected by multiple scattering, which lose the directional information. Thus, the optimal value of $\mathrm{w}$ for achieving the best $\mathrm{RMS}_{\theta}$ is determined by balancing these effects \cite{ghrear2025angularresolutionelectronsgaseous}. The values of $\mathrm{N}_{pt}$ and $\mathrm{w}$ that minimize the $\mathrm{RMS}_{\theta}$ of the distribution, after subtraction of the constant term, have been chosen for each energy and are presented in Tab.~\ref{tab:parametersvsE} for $w$, while fot $N_{pt}$ a constant value of 100 is used.
\begin{table}
    \centering
    \begin{tabular}{|l|cccccccccccc|}
    \hline
        E [keV] & 16 & 18 & 20  & 22 & 24 & 28 & 32 & 36 & 40 & 50 & 60 & 70 \\ \hline
        $\mathrm{w}$ & 1.5 & 1.5 & 1.5 & 1.5 & 1.5 & 1.5 & 2 & 2 & 2 & 2.5 & 2.5 & 2.5 \\ \hline
    \end{tabular}
    \caption{In the table, the optimal couples of parameters $\mathrm{N}_{pt}$ and $\mathrm{w}$ which provide the best $\mathrm{RMS}_{\theta}$ for each energy are shown.}
    \label{tab:parametersvsE}
\end{table}
For different energies, the value of $\mathrm{N}_{pt}$ remains constant. In contrast, the optimal value of $\mathrm{w}$ increases with rising energy levels. This trend occurs because, as energy increases, the effect of multiple scattering diminishes, making it more beneficial to include a larger portion of the track in the direction calculation.

\subsection{Directionality performance}
\label{sec:perform}
\begin{figure}
    \centering
    \includegraphics[width=0.75\linewidth]{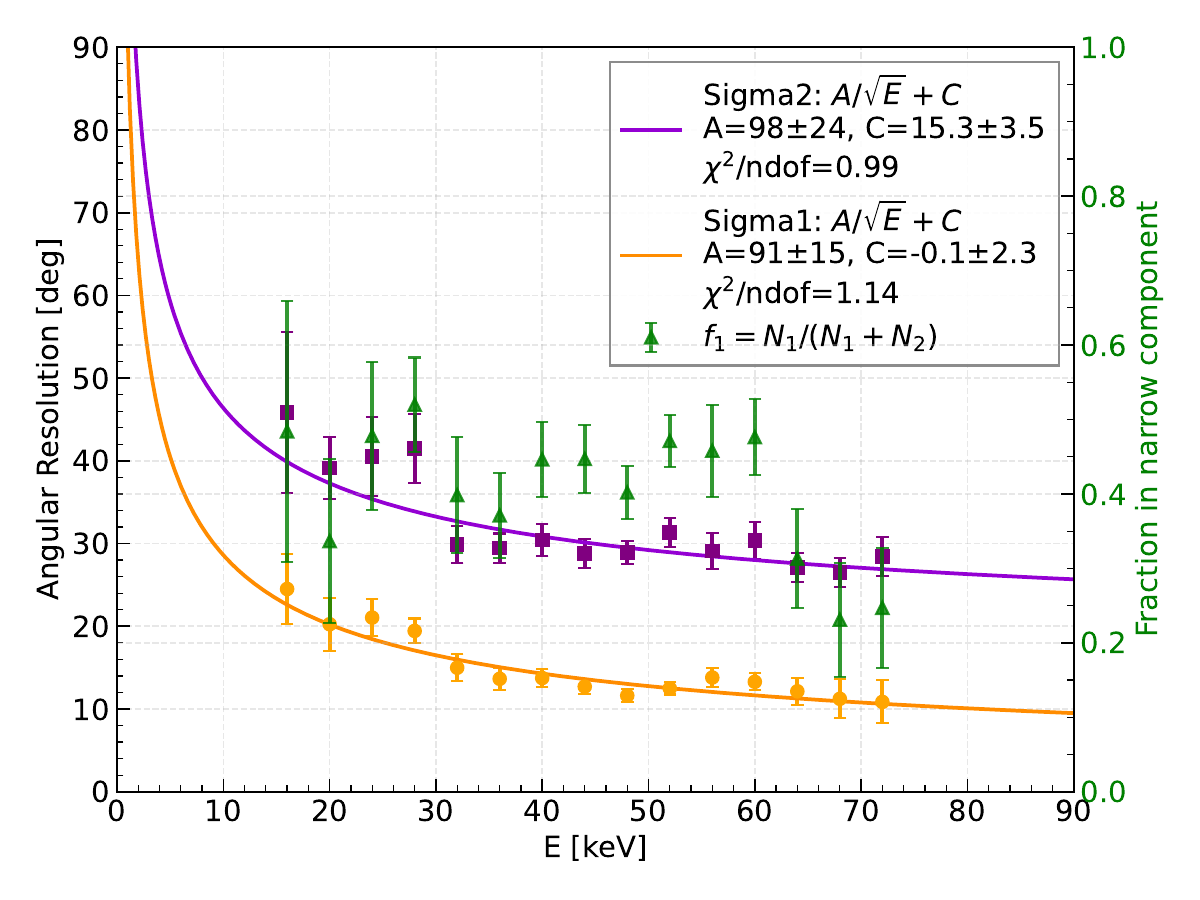}
    \caption{Angular resolution as a function of the track energy, expressed as the $\sigma$ of the narrower gaussian component (yellow) and of the broader gaussian component (purple), with the corresponding fit functions superimposed. In green also the fraction of events falling within the narrower gaussian over the sum of events falling in the two gaussians is shown. The curves are extended beyond the fitted energy interval
to illustrate the extrapolation of the empirical parametrization;
these regions are not constrained by the simulated data.}
    \label{fig:AngResIPResVsE}
\end{figure}

For each electron, the track energy in keV is determined from the light integral using the calibration curve of light response as a function of the energy, and the values of $\mathrm{N}_{pt}$ and $\mathrm{w}$ are assigned based on the closest energy reported in \ref{tab:parametersvsE}. The angular resolution is then evaluated starting form the sigmas of the two gaussians of the fit as a function of energy in bins. The plot illustrating the behavior of the angular resolution as a function of the energy is shown in Fig. \ref{fig:AngResIPResVsE}.
The angular resolution as a function of energy for the
two gaussian components has been fitted using the empirical
functional form $\sigma_{\theta}=A/\sqrt{E}+C$, where
$E$ is the reconstructed electron-recoil energy in keV,
and $A$ and $C$ are free fit parameters, with units of
degrees$\sqrt{\mathrm{keV}}$ and degrees, respectively.
The narrower (yellow) component characterizes the most
favourable projections, mainly associated with recoil
electrons emitted approximately parallel to the GEM plane.
For these tracks, the two-dimensional projection preserves
most of the initial directional information, resulting in
the best angular resolution. Its width can therefore be
considered an optimistic benchmark for the angular precision
of a future full three-dimensional reconstruction, in the
ideal limit where the track inclination is reconstructed
with negligible uncertainty.

The broader (purple) component describes less favourable
track inclinations, for which projection effects degrade
the angular resolution while some directional information
is still preserved. The constant term instead represents
events for which the reconstructed direction is effectively
random.

Since the angular response varies continuously with track
inclination, the two gaussian components provide an effective
description of the best- and worse-resolved directional
contributions rather than two distinct track populations.
The angular resolution was also investigated as a function
of drift distance, showing only a weak dependence over the
simulated range.

\begin{figure}
    \centering
    \includegraphics[width=0.75\linewidth]{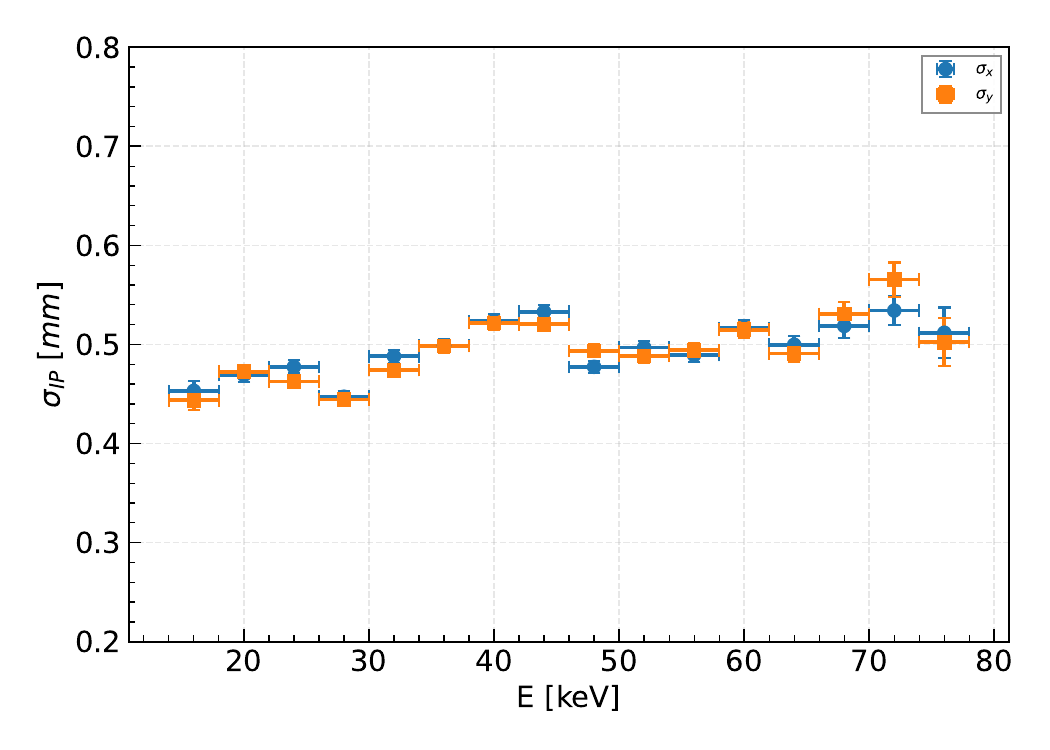}
    \caption{Spatial resolution of the reconstructed interaction point as a function of the incident energy, shown separately for the X (blue) and Y (orange) coordinates.}
    \label{fig:IPReso}
\end{figure}

Additionally, the interaction point resolution is assessed by fitting two gaussians to the distribution of the difference between the measured X and Y coordinates and those from the MC truth. The two gaussians accounts for the resolution and the effect of the efficiency in which a wrong interaction point is found. The interaction point resolution, shown in figure \ref{fig:IPReso}, maintain in the whole energy range below 570 $\mathrm{\mu}$m.

\subsubsection{Algorithm efficiency}
\label{sec:AlgoEff}
\begin{figure}
    \centering
    \includegraphics[width=.75\linewidth]{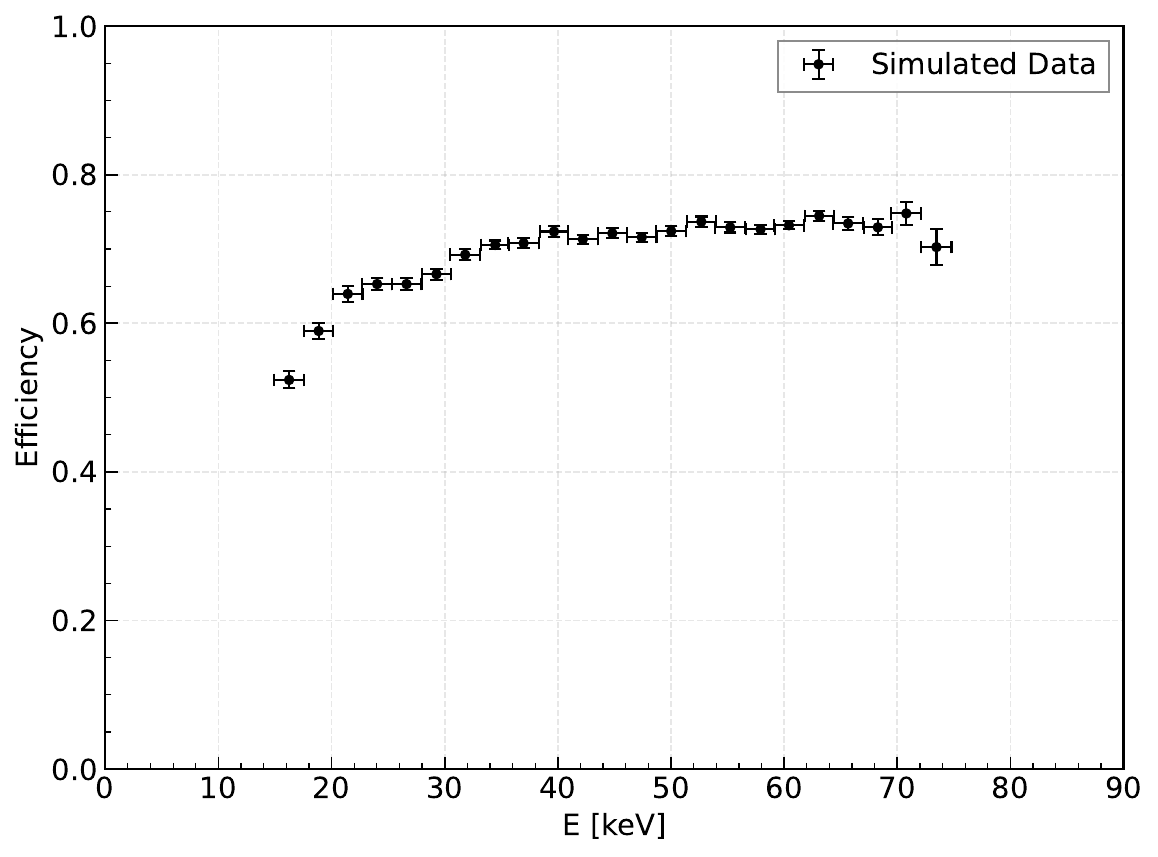}
    \caption{Directional efficiency, defined as the number of events falling within the gaussians fitting the $\mathrm{\theta}_{\mathrm{meas}}-\mathrm{\theta}_{\mathrm{true}}$ distribution divided by the total number of events, as a function of the electron energy. The horizontal error bars represent the bin widths.}
    \label{fig:efficiency}
\end{figure}

The cases in which the algorithm fails lead to incorrect identification of the interaction point, resulting in the reconstruction of a random track direction. As a consequence, a flat component appears in the $\mathrm{\theta}_{\mathrm{meas}}-\mathrm{\theta}_{\mathrm{true}}$ distribution, as shown in Fig. \ref{fig:gaussiandistrib}. The reconstruction efficiency of the algorithm is quantified as the ratio of the number of events falling into the gaussian components of the fit over the total number of events. Specifically, it is calculated as the integral of the gaussians of the fit divided by the total area under the fit function.
The situations in which the original polarimetry algorithm \cite{AstroXRayPol} fails to accurately identify the interaction point due to track topology were analyzed in \cite{LI201762}. These cases include instances where the track experiences a significant energy deposit in the middle due to hard scattering, which can affect the skewness value, or scenarios where the track's tail forms a curved arc, causing the algorithm to fail in selecting the correct track extremity.
In addition to these cases, there are situations where the track's initial direction is perpendicular to the GEM plane, resulting in the loss of the interaction point in the 2D track projection, making its identification possible only with PMT information.
The efficiency of directionality reconstruction as a function of energy is shown in Fig. \ref{fig:efficiency}. As expected, the efficiency improves with increasing energy due to longer tracks, which make the head and tail more distinguishable. Shorter tracks, particularly those produced perpendicularly to the GEM plane, present difficulties in retrieving the interaction point information. These shorter tracks are indeed the ones more susceptible to the effects of multiple scattering and diffusion, which affect the directionality calculation. As energy increases, the effects of multiple scattering and diffusion diminish, and the inefficiency becomes primarily driven by instances where the algorithm fails to identify the interaction point. Indeed, above a certain energy, the efficiency becomes energy-independent, leading to the convergence to a constant value.
Additionally, the efficiency has been analyzed for different drift distances and emission angles relative to the GEM plane. Since the analysis is performed on the 2D projection of the track, the efficiency of determining the initial track direction is notably affected when tracks are produced towards the GEM plane. However, no significant impact from diffusion is observed on the algorithm’s efficiency.\\
Integrating the third coordinate via PMT light sampling, alongside algorithmic improvement, is expected to enhance both the detection efficiency and angular resolution in the future.
The quoted uncertainties represent the statistical precision
of the simulated samples and the angular-response fits.
A recent simulation study has demonstrated the feasibility
of three-dimensional recoil-electron reconstruction by
combining two-dimensional optical imaging with waveform
information \cite{10.1093/ptep/ptag171}.
The detector-response model was previously validated against
experimental data \cite{Amaro:2026trz}.
Systematic uncertainties associated with variations of the
detector-response parameters have not been propagated
to the directional observables.
\section{Conclusions}
The capability to reconstruct the direction of low-energy electron recoils represents a key ingredient for directional measurements in rare-event searches, where directional information can provide powerful discrimination between signal and background and enable unambiguous signatures beyond spectral analyses alone. In this work, the directionality of low-energy electron recoils has been investigated in the context of the CYGNO experiment, focusing on a LIME-like optical time-projection chamber configuration optimized for high-granularity imaging of ionization tracks.
We have presented a dedicated reconstruction algorithm to infer the initial direction of electron recoils from two-dimensional optical images. The method is based on an approach originally developed for X-ray polarimetry and has been carefully adapted to handle the broader energy range and more complex track topologies characteristic of CYGNO-like detectors. By exploiting the intrinsic asymmetry of the energy-loss profile along the track, the interaction point is identified, and the direction is reconstructed by identifying the early portion of the ionization trail, where the information on the initial direction of the track is preserved.
The performance of the algorithm has been extensively studied using the detailed Monte Carlo simulations of a LIME, which include a realistic description of primary electron production, charge transport, amplification, light production, and sCMOS camera response. Following parameter optimization, the algorithm has demonstrated efficacy in determining the initial direction of electron recoils. The projected angular response is described by two Gaussian components that approximate its continuous dependence on track inclination. The narrower component, associated with favorable projections approximately parallel to the GEM plane, has a width of approximately \SI{20}{\degree} at \SI{20}{\kilo\electronvolt}, improving to about \SI{12}{\degree} at \SI{60}{\kilo\electronvolt}. The broader component, associated with projections more strongly affected by inclination, has a width of approximately \SI{40}{\degree} at \SI{20}{\kilo\electronvolt} and \SI{30}{\degree} at \SI{60}{\kilo\electronvolt}.
The narrower component therefore provides an optimistic
benchmark for the angular precision attainable with a
future three-dimensional reconstruction under favorable
conditions, while the broader component describes the
degradation associated with less favorable track projections.
Events with no significant directional information are
instead accounted for by the uniform component.
In addition, the reconstruction efficiency has been quantified, and the main sources of inefficiency have been identified and discussed. The directional reconstruction efficiency is approximately
50\% at 15 keV and increases to about 70\% at 30 keV. Beyond this point, the efficiency remains constant at $\sim$70\% with further increases in energy.
Overall, the results presented here demonstrate the performances of directional measurements of low-energy electron recoils in CYGNO-like optical TPCs and, as first result of this kind, provide a quantitative result of the achievable directional performance. The directionality performances obtained with LIME are expected to remain valid for larger CYGNO detectors, as they are designed to feature a readout granularity and a drift distance comparable to those of the LIME prototype. These results will serve also as a benchmark for future physics research exploiting directional electron-recoil detection, enabling realistic sensitivity studies for solar neutrino spectroscopy, dark matter searches, and other rare-event measurements where electron-recoil directionality provides a distinctive and powerful signature.

\acknowledgments
This project has received funding under the European Union’s Horizon 2020 research and innovation programme from the European Research Council (ERC), grant agreement No.~818744.
This project is supported by the Italian Ministry of Education, University and Research through PRIN grant 2020MZ884C, ``HypeX: High Yield Polarimetry Experiment in X-rays''.

\bibliographystyle{unsrt}
\bibliography{refs}

\end{document}